\documentclass[10pt,twocolumn,showpacs,superscriptaddress,preprintnumbers,floatfix]{revtex4}
\usepackage{latexsym}
\usepackage{amsmath}
\usepackage{amsfonts}

\usepackage{graphicx}

\def\laq{~\raise 0.4ex\hbox{$<$}\kern -0.8em\lower 0.62
ex\hbox{$\sim$}~}
\def\gaq{~\raise 0.4ex\hbox{$>$}\kern -0.7em\lower 0.62
ex\hbox{$\sim$}~}

\def\beq{\begin{equation}}
\def\eeq{\end{equation}}
\def\bea{\begin{eqnarray}}
\def\eea{\end{eqnarray}}
\def\bean{\begin{eqnarray*}}
\def\eean{\end{eqnarray*}}

\def\re {\rangle}

\def \a {\alpha}

   \def\be{\begin{equation}}
   \def\ee{\end{equation}}
   \def\ba{\begin{eqnarray}}
   \def\ea{\end{eqnarray}}

\begin{document}
\addtolength{\belowdisplayskip}{-0.0cm}
\addtolength{\abovedisplayskip}{-0.0cm}
\title{Backreaction during inflation: a physical gauge invariant formulation}
\author{F. Finelli}
\affiliation{INAF/IASF Bologna,
Istituto di Astrofisica Spaziale e Fisica
Cosmica di Bologna \\
via Gobetti 101, I-40129 Bologna - Italy}
\author{G. Marozzi}
\affiliation{Coll\`ege de France, 11 Place M. Berthelot, 75005 Paris, France and\\
\sl GR$\epsilon$CO - Institut d'Astrophysique de Paris, UMR7095 CNRS, \\ 
 Universit\'e Pierre \& Marie Curie, 98 bis boulevard Arago, 75014 Paris,
 France
}
\author{G. P. Vacca}
\affiliation{INFN, sezione di Bologna, via Irnerio 46, I-40126 Bologna, Italy}
\author{G. Venturi}
\affiliation{Dip. Fisica Universit\'a di Bologna and INFN, Sezione di Bologna, 
via Irnerio 46, I-40126 Bologna, Italy }
\begin{abstract}
Within a genuinely gauge invariant approach recently developed
for the computation of the cosmological backreaction, we study, in a
cosmological inflationary context and with respect to various
observers, the impact of {\it scalar} fluctuations on the space-time dynamics 
in the long wavelength limit. 
We stress that such a quantum backreaction effect is evaluated in a truly
gauge independent way using a set of effective equations which
describe the dynamics of the averaged geometry. 
In particular we show under what conditions the free falling (geodetic)
observers do not experience any
scalar-induced backreaction in the effective Hubble rate and fluid equation of
state. 
\end{abstract}
\maketitle
{\it Introduction}.
The computation of backreaction effects induced by cosmological fluctuations
in an inflationary era ~\cite{MAB1,ABM2}, has been the subject of controversial analysis
\cite{Uc,AW,GB,LU,FMVV_II}.
Such a task has been plagued by fundamental ambiguities
in constructing perturbatively gauge invariant (GI) 
observables ~\cite{Uc} and average quantities.
The basic fact that the averaging procedure does not commute with the non linear
evolution of Einstein equations~\cite{GE} was first exploited to study the
effective dynamics of the averaged geometry for a dust universe~\cite{Buchert}
(see~\cite{BO} for a recent application in the context of inhomogeneity driven inflation).
Gauge invariance of averaged quantities has recently been addressed in a novel context which
introduces a GI but observer dependent averaging prescription~\cite{GMV1}
(see~\cite{ULC} for a recent application of such a prescription to the analysis
of the present Hubble rate) whereas the effective equations for the averaged
geometry~\cite{Buchert} have been generalized in a covariant and GI form in~\cite{GMV2}.
Taking advantage of these recent results and having in mind the backreaction of quantum 
fluctuations during inflation, we devote this Letter to describing,
for the first time in this context,
an analysis of the GI effective equations which nevertheless depend on 
the different observers intrinsically used in the GI construction. 

{\it Gauge Invariant Backreaction}.
We start by illustrating how, following a recent proposal~\cite{GMV1,GMV2},
one may define observables, of a non local nature
and constructed with quantum averages,
which obey GI dynamical equations.
Specifically what has been investigated is how to give a classical or quantum
GI average of a scalar $S(x)$, for a classical field or a composite quantum operator,
assumed to be renormalized, respectively.
In such an approach the fundamental point is the choice of a hypersurface,
which defines a class of observers, with respect to which the average is done. 
In particular a hypersurface $\Sigma_{A_0}$ is defined, using another scalar
field $A(x)$ with a timelike gradient, through the constraint $A(x)=A_0$,
where $A_0$ is a constant. Both the scalars $S$ and $A$ are not GI
but one may construct an average which does have the desired property.
We simply consider here a spatially unbounded $\Sigma_{A_0}$ which is
reasonable in an inflationary context. 
Eventually one might constrain it to be bounded in the spatial
direction by using other scalar fields having a spacelike gradient, if available.
In investigations such as the actual dynamics of the universe,
one may imagine more suitable choices such as past light cone regions,
which however make the problem extremely more complicated. 

Once the GI definition is given, the computation of the
averaged quantity can be done in any gauge (coordinate frame).
We can define the quantum (classical, see notation of \cite{GMV1}) 
averaging prescription of a scalar quantity $S(x)$ as a functional of $A(x)$ 
which can be reduced,
in the (barred) coordinate system $\bar{x}^{\mu}=(\bar{t}, \vec{x})$ where the scalar
$A$ is homogeneous, to the simpler form~\cite{GMV1,GMV2}  
\beq
\langle S \rangle_{A_0}={\langle \sqrt{|\overline{\gamma}(t_0, {\vec{x}})|} 
\,~ \overline{S}(t_0, {\vec{x}}) \re \over  \langle   
\sqrt{|\overline{\gamma}(t_0, {\vec{x}})|} \re},
\label{media}
\eeq
where we have called $t_0$ the time $\bar{t}$ when $\bar{A}(\bar{x})=A^{(0)}(\bar{t})=A_0$.
The quantity $\sqrt{|\overline{\gamma}(t_0, {\vec{x}})|}$ is the square root of the
determinant of the induced three dimensional metric on $\Sigma_{A_0}$.
Note that in a general covariant definition the average is defined on a
spacetime region where the distribution $n^\mu\nabla_\mu\theta(A(x)\!-\!A_0)$ has support.
The vector $n^\mu$ defines the associated observer by $n^\mu= -Z_A^{-1/2}\partial^\mu A$,
$Z_A=-\partial^\mu A\partial_\mu A$ and $\theta$ is the Heaviside step function.  

Following the results of \cite{GMV1,GMV2} one can 
consider an effective scale factor $a_{eff}$ which describes the
dynamics of a perfect fluid-dominated early universe as
$a_{eff}=\langle\sqrt{|\bar{\gamma}|}\, \rangle ^{1/3}$ 
(where we have chosen $A^{(0)}(t)=t$ to 
have standard results at the homogeneous level \cite{Mar}),
and obtain a quantum gauge invariant version of the effective
cosmological equations for the averaged geometry:
\begin{widetext}
\bea
\left(\frac{1}{a_{eff}}\frac{\partial \, a_{eff}}{\partial A_0} \right)^2 \!\!&=&
\frac{8\pi G}{3} \left\langle\frac{\varepsilon}{Z_A}
\right\rangle_{\!\!\!A_0}\!\!
-\frac{1}{6}
\left\langle\frac{{\cal R}_s}{Z_A} \right\rangle_{\!\!\!A_0}\!\!
-\frac{1}{9}\left[   
\left\langle\frac{\Theta^2}{Z_A} \right\rangle_{\!\!\!A_0}\!\!\!
-\left\langle\frac{\Theta}{Z_A^{1/2}} 
\right\rangle_{\!\!\!A_0}^2 \right]\!\!
+\frac{1}{3} \left\langle\frac{\sigma^2}{Z_A} \right\rangle_{\!\!\!A_0}
=\, \,\frac{1}{9} \left\langle\frac{\Theta}{Z_A^{1/2}} 
\right\rangle_{A_0}^2\,,
\label{EQB1}\\
\!\!-\frac{1}{a_{eff}} \frac{\partial^2 \, a_{eff}}{\partial A_0^2}\!&=&\!
\frac{4 \pi G}{3} \left\langle\!\frac{\varepsilon+3 {\pi}}{Z_A} 
\!\right\rangle_{\!\!\!A_0}\!\!\!\!-\frac{1}{3}\left\langle
\frac{\nabla^\nu(n^\mu\nabla_\mu n_\nu)}{Z_A} \right\rangle_{\!\!\!A_0}
\!\!\!\!+\frac{1}{6} \left\langle\!\frac{n_\mu
\partial^\mu Z_A}{Z_A^{2}} \Theta \!\right\rangle_{\!\!\!A_0}
\!\!\!\!-\frac{2}{9}\!\left[   
\left\langle\!\frac{\Theta^2}{Z_A} \!\right\rangle_{\!\!\!A_0}\!\!\!
\!-\left\langle\!\frac{\Theta}{Z_A^{1/2}} 
\!\right\rangle_{\!\!\!A_0}^2 \right]\!\!
+\frac{2}{3} \left\langle\!\frac{\sigma^2}{Z_A}\! \right\rangle_{\!\!\!A_0}
\label{EQB2}
\eea
\end{widetext}
where ${\cal R}_s$ is a generalization of the intrinsic scalar 
curvature \cite{GMV2}, $\Theta=\nabla_\mu n^\mu$
the expansion scalar and $\sigma^2$ the shear scalar with respect to the
observer. We then define
\begin{eqnarray}
\varepsilon &=& \rho - (\rho+p)\left(1- (u^\mu n_\mu)^2 \right)\; , \
\\ 
\pi &=& p - \frac13 (\rho+p) \left(1- (u^\mu n_\mu)^2\right)\;,
\end{eqnarray}
with $u_\mu$ the 4-velocity comoving with the perfect fluid
and $\rho$ and $p$ are, respectively, the (scalar) energy density and 
pressure in the fluid's rest frame. In the inflationary scenario we consider
the inflaton field as the fluid.
Moreover we define the effective observer dependent energy density $\rho_{eff A}$
by writing the r.h.s. of Eq. \eqref{EQB1}
as $(8\pi G/3) \rho_{eff A}$ while the effective pressure $p_{eff A}$ 
is obtained by rewriting the r.h.s. of Eq. \eqref{EQB2} as
$(4\pi G/3) (\rho_{eff A}+3\, p_{eff A})$. 

In order to deal with the metric components in any specific frame we employ
the standard decomposition of the metric in terms of scalar, transverse vector 
($B_i$,$\chi_i$) and traceless transverse tensor ($h_{ij}$) 
fluctuations up to the second order
around a homogeneous FLRW zero order space-time
\bea
g_{00}\!\!&=&\!\! -1\!-\!2 \a\!-\!2 \a^{(2)}\,, \,\,  
g_{i0}=-{a\over2}\!\left(\beta_{,i}\!+\!B_i \right) \!
-\!{a\over2}\!\left(\!\beta^{(2)}_{,i}\!+\!B^{(2)}_i\!\right) 
\nonumber\\
\!\!\!\!g_{ij}\!\! &=&\!\!  a^2 \!\Bigl[ \delta_{ij} \! 
\left( \!1\!-\!2 \psi\!-\!2 \psi^{(2)}\right)
+D_{ij} (E+E^{(2)})
\nonumber\\
& & \!\!\!\!\!\!\!\!\!\!
+{1\over 2} \left(\chi_{i,j}+\chi_{j,i}+h_{ij}\right)
+ {1\over 2} \left(\chi^{(2)}_{i,j}+\chi^{(2)}_{j,i}+h^{(2)}_{ij}\right)\Bigr]\nonumber
\label{GeneralGauge}
\eea
where $D_{ij}=\partial_i\partial_j-\delta_{ij}\nabla^2/3$ and for notational
simplicity we have removed an upper script for first order quantities.
The Einstein equations connect those fluctuations with the matter ones. 
In particular the inflaton field is written to second order as
$\Phi(x)=\phi(t)+\varphi(x)+\varphi^{(2)}(x)$.

These general perturbed expressions can be gauge fixed. Let us recall
some common gauge fixing (of the scalar and vector part):
the synchronous gauge (SG) is defined by $g_{00}=-1$ and $g_{i0}=0$,
the uniform field gauge (UFG) apart from setting $\Phi(x)=\phi(t)$ must be
supplemented by other conditions (we consider $g_{i0}=0$),
finally the uniform curvature gauge (UCG) is defined by
$g_{ij}=a^2\left[\delta_{ij}+\frac{1}{2} \left(h_{ij}+h^{(2)}_{ij}\right)\right]$.

In the following we shall 
take the long wavelength (LW) limit as our approximation and
consider the cosmological backreaction with respect to different observers:

(a) the geodetic, or free falling, observers which are associated with a scalar
field homogeneous in the SG~\cite{Mar}. We consider such
observers to be the most interesting ones from a physical point of view.

(b) the observers associated with a scalar homogeneous in the UFG. We shall
show that up to second order in perturbation theory
they are equivalent to free falling observers.

(c) Let us also briefly comment on the possibility of defining an observer
which measures an unperturbed e-fold number $N$ in the LW limit,
i.e. an unperturbed effective expansion factor.
Consistently one finds, for such an observer which can be associated with a scalar
homogeneous in the UCG, identically zero backreaction effects in Eqs.
\eqref{EQB1} and \eqref{EQB2}.

{\it Geodetic Observers}.
We start by defining a free falling observer whose kinematics is determined by
the equation $t_\mu=v^\nu \nabla_\nu v_\mu=0$ for its velocity $v_\mu$,
which can be determined in any reference frame from the corresponding metric.
In our analysis we keep contributions up to second order in the fluctuations:
$v_\mu=v_\mu^{(0)}+ v_\mu^{(1)} +v_\mu^{(2)}$. In the SG one has $v_\mu=(-1,\vec{0})$.

Our first task is to define the scalar field $A(x)$ associated with this
observer.
Such a scalar field should give $n_\mu=v_\mu$ and appears to be the one 
homogeneous up to second order in the SG (see \cite{Mar} for the complete description).

For later use, let us exhibit the general condition for a scalar field $A(x)$ to be associated 
with free falling observers at first order. In this case $t_\mu$ should be zero
up to the first order.
The zero order condition is trivially satisfied for any scalar.
At first order the $\mu=0$ condition is always satisfied while the $\mu=i$ condition gives
\be 
\frac{d}{dt}\left(\frac{A^{(1)}}{\dot{A}^{(0)}}\right)-\alpha=0\,.
\label{Cond_Geodesic}
\ee
As is easy to check, the l.h.s. of this condition is GI since the
vector $t_\mu^{(0)}$ is identically zero. 

In general, using the coordinate transformations up to second order \cite{MetAll}
\be
x^\mu \to \bar{x}^\mu=x^\mu+\epsilon^\mu_{(1)}+
\frac{1}{2} \left( \epsilon^\nu_{(1)}\partial_\nu \epsilon^\mu_{(1)}+\epsilon^\mu_{(2)}\right)\,,
\label{CoordTrasf}
\ee
we
define the espressions associated with Eq.~\eqref{media}, going from a
general coordinate system to the barred one.

Our strategy is therefore the following: 
we shall construct the observables and study Eqs.~(\ref{EQB1},\ref{EQB2}) using 
Eq.(\ref{media}). 
The results are by definition GI, this allows us to use the results for the
dynamics of the inflaton and metric fluctuations, which satisfy the Einstein
equations, up to second order, in any frame convenient for the calculations 
(we shall use the results computed in the UCG~\cite{FMVV_II}).

In the LW limit we have
\begin{eqnarray}
& & \bar{\Theta}=3H-3 H \bar{\alpha}-3\dot{\bar{\psi}}
+\frac{9}{2} H \bar{\alpha}^2+3 \bar{\alpha} \dot{\bar{\psi}}
-6\bar{\psi}\dot{\bar {\psi}}
\nonumber \\
& & \,\,\,\,\,\,\,\,\,\,\,\,\,-3 H \bar{\alpha}^{(2)}-
3 \dot{\bar{\psi}}^{(2)}-\frac{1}{8}h_{ij}\dot{h}^{ij}
\label{Theta} \\
& & \!\!\!\!\!- \partial_\mu \bar{A} \partial^\mu \bar{A}= 1- 2 \bar{\alpha}+4 
\bar{\alpha}^2- 2 \bar{\alpha}^{(2)}
\label{partA}
\end{eqnarray}
and for the measure in the spatial section
\be 
\sqrt{|\bar{\gamma}|}=a^3 \left(1-3 \bar{\psi}+\frac{3}{2}\bar{\psi}^2
-\frac{1}{16} h^{ij}h_{ij}
-3\bar{\psi}^{(2)}\right)\,.
\label{detgamma}
\ee 
Let us note that we shall neglect in our computations the dependence on the tensor fluctuations
hereafter (see however \cite{FMVV_GW} for the backreaction 
of tensor fluctuations in de Sitter space-time). 
Inserting Eqs.(\ref{Theta}, \ref{partA}, \ref{detgamma}) in
Eq.(\ref{EQB1})
one obtains the simple expression
\be
\left(\frac{1}{a_{eff}}\frac{\partial \, a_{eff}}{\partial A_0} \right)^2 =
H^2 \left[1+\frac{2}{H}\langle \bar{\psi}\dot{\bar{\psi}}\rangle-
\frac{2}{H}\langle \dot{\bar {\psi}}^{(2)}\rangle\right]
\label{EQ1simpl_2}
\ee

Let us turn our attention to the SG observers.
The coordinate transformations needed to go to the SG are characterized by
\be
\epsilon_{(1)}^0=\! \int^t \!\!dt' \alpha \,\,\,,\,\,\, 
\epsilon_{(2)}^0= - \alpha \!\int^t\!\! dt' \alpha + \int^t \!\!dt' \left( 2 \alpha^{(2)} \!-\!
\alpha^2 \right) \,, \nonumber
\ee
where we neglect a non dynamical constant contribution.
Using Eq.(\ref{CoordTrasf}) we have
\begin{eqnarray}
& &\bar{\alpha}=0\,\,\,\,,\,\,\,\,\bar{\psi}=\psi+H \int^t dt' \alpha\,\,\,\,,\,\,\,\,\bar{\varphi}=\varphi
-\dot{\phi} \int^t dt' \alpha \nonumber \\
& & \bar{\psi}^{(2)}=\psi^{(2)}-H \alpha \int^t dt' \alpha-\frac{1}{2} \left(\dot{H}+2 H^2\right)
\left[\int^t dt' \alpha\right]^2
\nonumber \\
& &-\left(2 H \psi+\dot{\psi}\right)\int^t dt' \alpha+
\frac{H}{2} \int^t dt' \left(2 \alpha^{(2)}-\alpha^2\right) \nonumber
\end{eqnarray} 
In order to evaluate the backreaction Eq. (\ref{EQB1}), as said, we choose to perform the calculation 
in the UCG.
In this gauge we need the solution to the equations of motion for the inflaton and the metric.
To first order, one has
\be
\alpha=\frac{1}{2M^2_{pl}}\frac{\dot{\phi}}{H} \varphi \,,
\label{Eq_usefull1}
\ee
where $M_{pl}^{-2}=8\pi G$,
and considering only the LW limit one gets
\be
\varphi=f(\vec{x}) \frac{\dot{\phi}}{H}  \Rightarrow
\int^t \!dt' \alpha=-f(\vec{x}) \int^t \! dt'
\frac{\dot{H}}{H^2}=\frac{1}{\dot{\phi}} \varphi\,. 
\label{Eq_usefull2}
\ee
These lead to
$\bar{\varphi}=0$, $\bar{\alpha}=0$ and
$\bar{\psi}=\frac{H}{\dot{\phi}} \varphi$
where the last term, which we insert in~\eqref{EQ1simpl_2}, is constant in such a limit.

To second order we restrict to the particular case of a non self-interacting massive 
inflaton field~\cite{FMVV_II}. In such a case, in the LW limit and at the
leading order in the slow-roll approximation, one has

\begin{eqnarray}
\langle \alpha^{(2)} \rangle = \frac{1}{M_{pl}^2}\,\epsilon \,
 \langle \varphi^2 \rangle \,, \nonumber
\end{eqnarray}
where $\epsilon=-\dot H/H^2$.
We then obtain
\be
\langle\dot{\bar{\psi}}^{(2)}\rangle= H \,{\cal O}\left(\epsilon^2\right)
 \frac{\langle \varphi^2 \rangle}{M_{pl}^2}\,.
 \nonumber
\ee
Inserting the various results into Eq. \eqref{EQ1simpl_2}, we find
\be
\left(\frac{1}{a_{eff}}\frac{\partial \, a_{eff}}{\partial A_0}
\right)^2=H^2\left[1+ {\cal O} \left(\epsilon^2 \right) \frac{\langle \varphi^2 
\rangle}{M_{pl}^2}\right]\,,
\label{reseq1}
\ee
where, for a massive chaotic model~\cite{FMVV_II} in the LW limit and $H_i=H(t_i)\gg H$, one has
\be
 \frac{\langle \varphi^2\rangle}{M_{pl}^2}\simeq-\frac{1}{24\pi^2} 
\frac{H^6_i}{M_{pl}^2 H^2 \dot{H}}  \sim \frac{H_i^4}{H^2 M_{pl}^2} \ln a
\label{phiquad}
\ee
(see~\cite{starall} for a generic single field inflationary scenario).
Therefore there is no {\it leading} backreaction in the slow-roll parameter $\epsilon$
on the effective Hubble factor induced by scalar fluctuations.
Provided the coefficient of $\langle \varphi^2 \rangle$ in
Eq.~\eqref{reseq1} does not turn out to be zero,
the quantum backreaction has the chance of appearing in the next-to-leading
order, with a secular term related to the infrared growth of inflaton fluctuations.
On the other hand such a growth gives a negligible effect whenever the quantity in
\eqref{phiquad} is much smaller than $\epsilon^{-2}$.
One can proceed in a similar way with the analysis of the second backreaction 
equation (\ref{EQB2}) by evaluating directly the expressions. However we give here
a more general result valid for any observer and slow-roll inflationary
models.
If for the effective Hubble factor one finds 
\be
\left(\frac{1}{a_{eff}}\frac{\partial \, a_{eff}}{\partial A_0}
\right)^2\!=H^2\!\left[1\!+ \left( c \,\epsilon^n\! +\!{\cal O} \left(\epsilon^{n+1}\right)\right)
\frac{\langle \varphi^2 \rangle}{M_{pl}^2}\right]\,,
\nonumber
\ee
then, from the consistency between the effective equations for the averaged
geometry, one obtains
\be
-\frac{1}{a_{eff}} \frac{\partial^2 \, a_{eff}}{\partial A_0^2}=   
-\dot{H}\!-\!H^2\!-\!H^2 \!\left[c \, \epsilon^n \!+\!
{\cal O} \left(\epsilon^{n+1}\right)\right]
\frac{\langle \varphi^2 \rangle}{M_{pl}^2}\,
\nonumber
\ee
and it is easy to see that the effective equation of state is unperturbed up to the leading
non trivial order, i.e.
\be
w_{effA}=\frac{p_{effA}}{\rho_{effA}}=-1+\frac{2}{3} \epsilon
+{\cal O} \left(\epsilon^{n+1} \right)
\frac{\langle \varphi^2 \rangle}{M_{pl}^2}\,.
\nonumber
\ee

For the SG observer one has $n=2$ and we obtain the same condition as before
to have no appreciable scalar backreaction effects.
Non negligible effects could appear at the end of inflation 
($H\sim m$) for initial conditions such that $H(t_i)\sim (m^2 M_{pl})^{1/3}$,
which is an energy scale much smaller than the Planck one but is
however associated with an extremely long inflationary era.
Indeed such values give a typical number of e-folds of the order of ${\cal O}(10^4)$, 
for $M_{pl}=10^5m$, and  correspond to the case where non-linear corrections become
really important~\cite{starall,BPT}. In this case other computational
techniques are required.

{\it UFG Observer}.
Let us introduce the observers associated with a scalar homogeneous in the UFG.
Such an observer always sees as homogeneous the inflaton field $\Phi$
along its whole evolution during the inflationary regime.
It is easy, following \cite{Mar}, 
to identify the scalar associated to this observer as
\be
A(x)=A^{(0)}+\frac{\dot{A}^{(0)}}{\dot{\phi}} \varphi 
+\frac{\dot{A}^{(0)}}{\dot{\phi}} \varphi^{(2)}
+\frac{\dot{A}^{(0)}}{2 \dot{\phi}^{2}} \left(
\frac{\ddot{A}^{(0)}}{\dot{A}^{(0)}}-
\frac{\ddot{\phi}}{\dot{\phi}} \right)\varphi^{2} \,.
\nonumber
\ee
In this way the condition (\ref{Cond_Geodesic}) to have geodesic observers to
first order becomes
\be 
\frac{d}{dt}\left(\frac{\varphi}{\dot{\phi}}\right)-\alpha=0\,.
\label{Cond_Geodesic_UFG}
\ee
Such a condition is trivially satisfied in the LW limit (as can be seen in the
UCG using Eqs.(\ref{Eq_usefull1},\ref{Eq_usefull2})).
To see if this equivalence is valid also to second order one should study the 
condition $t_\mu^{(2)}=0$.
The quantity $t_\mu^{(2)}$ is GI in this case since
$t_\mu^{(0)}=t_\mu^{(1)}=0$, and the condition can be studied in any gauge.
Choosing the barred gauge it becomes $\bar{\alpha}^{(2)}=0$ and is satisfied
in the LW limit.
As a consequence the UFG observers are physically equivalent to the free falling ones 
and experience the same backreaction, as can be explicity verified by calculations.

Let us emphasize that this property is not valid in general for all observers in the LW limit.
Let us, for example, consider the observer define by the scalar homogeneous 
in the longitudinal gauge 
($\beta=E=0$).
This is defined to first order by
\be
A(x)=A^{(0)}+\dot{A}^{(0)} \left[\frac{a}{2}\beta-\frac{a^2}{2}\dot{E}\right]
\nonumber
\ee
and the condition (\ref{Cond_Geodesic}) is not verified as can be easily seen. 
Such observer is not in geodesic motion and 
it may see a backreaction effect, even to
leading order in the slow-roll parameter, which is in general different
with respect to that experienced by a free falling observer.
This case will be studied elsewhere.

{\it Conclusions}.
We have applied for the first time the gauge invariant observer dependent
approach introduced in~\cite{GMV1,GMV2}
to analyze backreaction effects induced by long wavelength scalar fluctuations in the
cosmological early universe during an inflationary era.

We have seen how for geodetic observers the backreaction of 
long wavelength scalar fluctuations does not appear to leading order 
in the slow-roll approximation for a $m^2 \phi^2$ chaotic
inflationary model.  
Moreover we have shown under what conditions the backreation is negligible or
not in the next-to-leading order. 
In the same long wavelength limit the  tensorial contribution also disappear.
This is a physical result which is derived in a truly gauge invariant way.

Modes with shorter wavelengths, which typically behave less classically, 
may be a source of backreaction seen by physical observers,
through all the terms, present in the equations, which have been neglected here.
Second order tensor and scalar contributions, the latter depending also on
the former, could also be investigated within the present framework. 
Computations are much more involved but will give a consistent gauge invariant 
answer to such a question, at least for the second order perturbative expansion.

\end{document}